\documentclass[sn-nature]{sn-jnl}

\usepackage{graphicx}%
\usepackage{multirow}%
\usepackage{amsmath,amssymb,amsfonts}%
\usepackage{amsthm}%
\usepackage{mathrsfs}%
\usepackage[title]{appendix}%
\usepackage{xcolor}%
\usepackage{textcomp}%
\usepackage{manyfoot}%
\usepackage{booktabs}%
\usepackage{algorithm}%
\usepackage{algorithmicx}%
\usepackage{algpseudocode}%
\usepackage{listings}%
\theoremstyle{thmstyleone}%
\theoremstyle{thmstyletwo}%

\theoremstyle{thmstylethree}%

\begin{document}

\title[]{Floating binary planets from ejections during close stellar encounters}


\author*[1,2]{\fnm{Yihan} \sur{Wang}}\email{yihan.wang@unlv.edu}
\equalcont{These authors contributed equally to this work.}
\author[3,4]{\fnm{Rosalba} \sur{Perna}}\email{rosalba.perna@stonybrook.edu}
\equalcont{These authors contributed equally to this work.}
\author[1,2]{\fnm{Zhaohuan} \sur{Zhu}}\email{zhaohuan.zhu@unlv.edu}

\affil*[1]{\orgdiv{Nevada Center for Astrophysics}, \orgname{University of Nevada}, \orgaddress{\street{ 4505 S. Maryland Pkwy.}, \city{Las Vegas}, \postcode{89154}, \state{NV}, \country{USA}}}

\affil[2]{\orgdiv{Department of Physics and Astronomy}, \orgname{University of Nevada}, \orgaddress{\street{505 S. Maryland Pkwy.}, \city{Las Vegas}, \postcode{89154}, \state{NV}, \country{USA}}}

\affil[3]{\orgdiv{Department of Physics and Astronomy}, \orgname{Stony Brook University}, \orgaddress{\street{100 Nichols road}, \city{Stony Brook}, \postcode{11794}, \state{NY}, \country{USA}}}

\affil[4]{\orgdiv{Center for Computational Astrophysics}, \orgname{Flatiron Institute}, \orgaddress{\street{162 5Th Ave}, \city{New York}, \postcode{10010}, \state{NY}, \country{USA}}}


\abstract{
 The discovery of planetary systems beyond our solar system has
challenged established theories of planetary formation.  Planetary
orbits display a variety of unexpected architectures, and
free-floating planets appear ubiquitous.  The recent detection of
candidate Jupiter Mass Binary Objects (JuMBOs) by the {\em James Webb Space
Telescope} (JWST) has added another puzzling layer. Here, through
direct few-body simulations, we demonstrate that JuMBOs could arise
from the ejection of double giant planets following a close encounter
with a passing star, if the two planets are nearly aligned at closest
approach.  These ejected JuMBOs typically possess an average
semi-major axis approximately three times the orbital separation
within their original planetary system and a high eccentricity,
characterized by a superthermal distribution that sets them apart from
those formed primordially.  We estimate the JuMBO formation rate per
planetary system in typical and densely populated clusters, revealing
a significant environmental dependence. In dense clusters, this
formation rate can reach { a few} percent for wide planetary
systems.  Comparative analysis of JuMBO rates and properties with
current and forthcoming JWST observations across various environments
promises insights into the conditions under which these giant planets
formed in protoplanetary disks, thereby imposing constraints on
theories of giant planet formation.
}




\maketitle

The discovery and investigation of over {5,000} exoplanets { \citep{Christiansen2022}}
beyond our
Solar System  has unveiled a remarkable variety of exoplanets and shown that our own planetary
system is far from typical \citep{ZhuDong2021}. 
{Particularly, the discovered giant planets have turned out to be the most puzzling population, challenging the conventional theory of giant planet formation.}

{ In the conventional core accretion theory, a solid core is first assembled through planetesimal accretion \citep{Pollack1996I} or pebble accretion \citep{OrmelKlahr2010,LambrechtsJohansen2012}. As the core's gravity becomes strong, a
hydrogen-helium gaseous envelope starts to develop around the planetary core. This envelope
accretion phase is the longest stage of giant planet formation due to the envelope's slow Kelvin-Helmholtz contraction. If the envelope mass could eventually reach the core mass within the disk lifetime, the envelope would accrete exponentially, and the planet would enter the runaway stage to become a full-grown giant planet. With nominal disk parameters, the core mass needs to be $\gtrsim 10M_{\oplus}$ for runaway to occur before the gaseous disk dissipates within about $\sim$3 Myrs. The ice particles beyond the frost line help supply additional solids to build this  massive core.}

{ Although this conventional model could explain our solar system, it meets some challenges in explaining giant exoplanets.
As a start, the `hot Jupiters', the first discovered exoplanets around main-sequence stars \citep{MayorQueloz1995}, are found to be situated remarkably close to
their host stars, well within the frost line. Second, many of these `hot Jupiters' inhabit highly eccentric orbits and display significant relative inclinations \citep{WinnFabrycky2015},  at odds with the circular co-planar orbits of giant planets in our own solar system.  } Third, the recent detection of giant planets orbiting low-mass stars \cite{Morales2019} suggests efficient giant planet formation, defying 
the core accretion theory. Fourth, this theory is further questioned by the observation of brown dwarfs/planets on extremely wide orbits, larger than $\sim 100$~AU \citep{Marois2008}, 
{approaching $\sim 10^3$~AU
\cite{Lafreniere2011}, and possibly exceeding it \citep{Deacon2016}}. 
Finally,
the presence of floating planets has also been another puzzle for this theory \citep{Miret-Roig2022}.
Very recently, the mystery has deepened with the discovery by the {\em James Webb Space Telescope (JWST)} \cite{Pearson2023}
that a fraction of these floating planets, largely Jupiter-like giants, is moving in a couple, thus making up a new population of binary planets whose existence does
not readily fit in any current planetary formation theory.
{ Although some of these challenges can be remedied by invoking additional physical processes (e.g., planet migration \citep{Lin1996}, planet-star tidal interaction \citep{Winn2010}, planet-planet scattering { \citep{Chatterjee2008,Miret-Roig2022}}, more realistic disk physics \citep{Lee2019}), most of these processes could only operate for planets whose semi-major axes are within several AU. It is still challenging to explain giant planets far away from the star \citep{LambrechtsJohansen2014}, and the abundant free-floating planets, using the core accretion theory \citep{Veras2012}.

{ On the other hand, the alternative theory of giant planet formation, the disk gravitational instability model \cite{Boss1997}, can form giant planets efficiently beyond 50~AU. When young protoplanetary disks are massive enough, they are subject to gravitational instability and develop spiral arms. If disk cooling is fast enough, these spirals can fragment and collapse to giant planets directly \citep{Gammie2001}. With typical protoplanetary disk conditions, this fast cooling can be achieved beyond 50 AU \citep{Rafikov2005}. The clumps collapse quickly and  form planets with several Jupiter masses. Although this theory has its own challenges, particularly in explaining terrestrial planets close to the star (but see \cite{Boley2010,Nayakshin2011}), it can form multiple giant planets efficiently beyond 50~AU at early disk evolutionary stages \citep{Zhu2012}.}

{It is difficult to test these two formation models using mature exoplanets that are billions of years old. It is crucial to discover young giant planets. Unfortunately, we only have a few candidates discovered in protoplanetary disks (e.g. PDS 70bc at 21 and 34~AU from the central star \citep{Keppler2018,Haffert2019}).
Intriguingly, a promising avenue of exploration has emerged that ventures beyond intrinsic mechanisms—namely, the influence of external perturbations on planetary architectures (e.g. \cite{Laughlin1998,Bonnell2001,Ford2006,Malmberg2007,fregeau06,Malmberg2011,Hao2013,Shara2016,Cai2017,Flammini2019,Fragione2019a,Li2019,Wang2020a,Wang2020b,Li2020,Moore2020,Wang2020c,Carter2023, Chen2024}). Planetary systems are in fact likely born in young star clusters. 
In such dense stellar environments, frequent gravitational interactions between celestial objects become commonplace, potentially reshaping planetary systems over time. This dynamic interplay in crowded clusters offers an alternative perspective on planetary formation and the planet orbital configurations at early formation times, addressing some of the existing gaps in our understanding.

This work has been {  inspired} by the very recent report  of 
{\em JWST} observations  of candidate Jupiter-Mass Binary Objects (JuMBOs) which are also quite young \cite{Pearson2023}.   
More specifically, via dedicated $N$-body simulations, we set to
investigate the possibility that a close flyby can result in the ejection
of two planets in outer orbits, which then remain bound to one another. 
While the initial motivation for this exploration came from the observations mentioned above, our simulations robustly demonstrate that JuMBO formation is actually an { unavoidable} outcome of close-by interactions in dense stellar environments, and the properties of JuMBOs can be informative of their initial configurations in protoplanetary disks. 
We quantify their occurrence as a function of the type of stellar environment and the original planetary configuration and make predictions for their properties which can be tested with future {JWST} observations.

\subsection*{Results}\label{sec2}

Our planetary system consists of two equal-mass planets with mass $m$ in circular orbits with semi-major axes $a_1$ and $a_2$, and a host star with mass $M_1$.
Both planets move in circular
orbits around the host star with 
initial phases $\nu_1$ and $\nu_2$ with respect to the $X$ axis, and velocities $v_1$ and $v_2$ 
corresponding to the orbital semi-major axes (SMA) $a_1$ and $a_2$, respectively.
The flyby intruder with mass $M_2$ approaches the planetary system with impact parameter $b$ and asymptotic velocity at infinity $v_\infty$.  We assume the orientation of the planetary system to be isotropically
distributed, which translates into the parameter distributions
$\cos I\in [-1,1]$ and $\Omega\in[0,2\pi]$, where $I$ is the orbital inclination, and $\Omega$ is the longitude of the ascending node. The schematics of the scattering experiment, with all the variables involved, are shown in Fig.~\ref{fig:schematics1}.
Throughout the paper, we adopt a generic (fixed) mass ratio $m/M_1=10^{-3}$ (similar to the mass ratio between Jupiter and the Sun), 
and $M_2/M_1=1$ for simplicity. We ran $10^8$ scattering experiments 
for each combination of $(v_\infty/v_2, a_1/a_2)$ and found that JuMBO form as an outcome of ejection following the flyby, especially if the two giant planets are nearly aligned when the intruder is at its closest approach. We explore and quantify their properties in the following.

We begin by investigating the dependence of the relevant JuMBO properties on the angles $I$ and $\Omega$. This is shown in Fig.~\ref{fig:Omega-I} for a fixed value $a_1/a_2=0.7$ and three choices of $v_\infty/v_2$. From top to bottom, we show the relative cross section (normalized to the maximum value) of JuMBO formation, the semi-major axis (SMA) of the JuMBO normalized to $\Delta a = (a_2-a_1)$, and the eccentricity. A notable characteristic of the cross section for JuMBO formation is the highest probability for edge-on scatterings when either $\Omega$ or $I$ is $0$ or $\pi$ (symmetry exists with respect to these values as $\theta\in[0,2\pi]$). In contrast, the probability is lowest for face-on scatterings where $\Omega=I=\pi/2$. During face-on scatterings, the interloper interacts with the planetary system for a relatively short duration as it crosses the planetary orbital plane. However, in edge-on scatterings, the interloper interacts with the planetary system for a much longer time as it travels along the orbital plane. This is especially true for prograde edge-on scatterings, where the interloper and the two planets move in the same direction. Their minimal relative velocity increases the interacting time, leading to a higher chance of ejecting two giant planets consecutively.

The semi-major axis (SMA) of a JuMBO is largest when the scattering is close to face-on and smallest when the scattering is edge-on. JuMBOs tend to form when the two giant planets are nearly aligned as the intruder approaches its closest point. At this time, their separation upon ejection is approximately equal to the difference in their SMA, and their relative velocity difference aligns closely with their Kepler velocity difference. In edge-on scatterings, the intruder imparts momentum parallel to the planets' velocities. The outer planet, which has a lower Keplerian velocity, gains more momentum since it is closer to the intruder. In contrast, the inner planet gains relatively less momentum due to its greater distance from the intruder. Consequently, the intruder tends to decrease their relative velocity upon ejection. For face-on scatterings, the imparted momentum is perpendicular to the Keplerian velocities of the two planets. This results in a weaker relative velocity reduction compared to edge-on scatterings. A smaller relative velocity upon ejection results in a smaller binding energy post ejection, leading to a decreased SMA as indicated by Eq~\ref{eq:a}. Due to the reduction in relative velocity, the eccentricity of the JuMBOs tends to be very high, stemming from the small angular momentum between the two ejected planets. Edge-on scatterings are more efficient at reducing this relative velocity, which is why we observe JuMBOs with higher eccentricities resulting from these scatterings.
{ Note that we find that the events causing the JuMBO ejections are dominated by impact parameters
 $b\lesssim 24~a_2$ for $v_\infty/v_2=0.1$, by  $b\lesssim 2.6~a_2$ for $v_\infty/v_2=1$ and $b\lesssim 0.5~a_2$ for $v_\infty/v_2=10$.}

Fig.~\ref{fig:cross-sections}
compares the cross sections for ejection of a single planet
(top panel) with that of JuMBOs (bottom left panel). As expected, both are favored by lower relative velocities of the flyby star. However, while the ejection of a single planet is largely independent of the relative initial orbits of the two planets, the probability of ejection of JuMBOs increases as the two planets are initially in more closely spaced orbits. This becomes more apparent in the ratio between the cross sections, which is shown in the bottom right panel of  Fig.~\ref{fig:cross-sections}.   

To relate our dimensional outcomes to direct astrophysical environments, the figures explicitly indicate with vertical lines the location of the ratio $v_\infty/v_2$ for an outer planet at $a_2=10$~AU (blue lines) and at $a_2=100$~AU (red lines), each for three values of the velocity dispersion, $\sigma_v=1,5,10$~km/s, roughly corresponding to the typical values encountered in open clusters, globular clusters, and OB associations, respectively. 

The orbital properties of the dynamically-formed JuMBOs are displayed in Fig.~\ref{fig:jumbo-ae}, with the top panels showing the mean and standard deviation (STD) of the SMA, and the bottom panels displaying the same quantities for the eccentricity. The general
features of this JuMBO population are a large eccentricity, mostly in the range of $e_{\rm JuMBO}\sim 0.65-0.75$, with a standard deviation $\sim 0.3$. The mean of the semi-major axis $a_{\rm JuMBO}$
has a spread of a few around $\Delta a$, with an STD also of a few. As intuitively expected, tighter binaries are formed for initially closer separations between the two planets. 

Last, we compute and show in Fig.~\ref{fig:dist-ae} the probability density function (PDE) for the SMA (upper panel) and the eccentricity (bottom panel) of the JuMBOs, for four representative values of $v_\infty/v_2$ and the value of 
$a_1/a_2=0.7$. These PDEs hence represent the distributions that observations of JuMBOs formed via ejection would be seeing. It has been demonstrated that in the fast scattering regime, where  $v_\infty/v_2\gg 1$, the SMA distributions exhibit distinctive shapes \citep{heggie75,Hut83,sigurdsson93}. Since the two planets in a JuMBO are nearly aligned upon ejection, the intruder imparts less momentum to the planets in the fast scattering regime than in the slow scattering regime. This is attributed to the relatively shorter interaction time in the fast scattering regime. As a result, the reduction in relative velocity as discussed above is less efficient in a fast scattering interaction. Consequently, achieving a smaller SMA for the JuMBO is more challenging in the fast scattering regime. 

We further observe that, while the PDE of $a_{\rm JuMBO}$ is peaking around the value {$\sim3$}$(a_2-a_1)$, it  has however a broad tail at larger values. Hence forming JuMBOs with large SMA is natural within the mechanism of our study. Planets at very large separation are known to exist \cite{Lafreniere2011}, while more typical outer planets at tens of AU can easily form JuMBO with SMA in the 100s~AU range. However, if a large sample of observations shows that the JuMBO distribution is very heavily dominated by wide binaries, with $a_{\rm JuMBO}\gtrsim 100$~AU (as suggested by the first set of JuMBO candidates \cite{Pearson2023}), then the JuMBO formation mechanism identified in this work would support the giant planet formation models that can efficiently produce wide-orbit giant planets in young protoplanetary disks, e.g.
the disk gravitational instability model \cite{Boss1997}. Meanwhile, the two planet formation models are not mutually exclusive, and the giant planet population at $\lesssim$10~AU could still form through core accretion at later disk evolutionary stages. 

The JuMBO eccentricity distribution 
is found to be superthermal, as can be seen in the bottom panel of Fig.~\ref{fig:dist-ae} as compared to the thermal distribution. This is a very distinctive feature of our JuMBO formation mechanism since the eccentricity distribution expected in scenarios of primordial formation from Interstellar Medium (ISM) clouds is expected to be subthermal \citep{Mathieu1994,raghavan10}.

The absolute rate of JuMBO formation is dependent on the stellar density in the interacting cluster, which, in virialized systems, can be determined from their mass and velocity dispersion. However, it is also dependent on the highly uncertain fraction of planetary systems with giant planets in outer orbits, which hence makes a numerical evaluation of the absolute rate  
{rather approximate}
at this stage. 
Nonetheless, using the formalism of rate calculation detailed in Sec.4.2 of the Methods, we can make an estimate for the number of JuMBOs produced via our proposed mechanism. In particular, in the optimal scenario where every planetary system contains at least two giant planets, we can derive an upper limit for the JuMBO production efficiency per planetary system, as a function of $a_1$ and $a_2$. 

The upper panel of Fig~\ref{fig:rate} displays the magnitude of this
upper limit for the specific case of the inner Trapezium region of the Orion Nebula Cluster (ONC) over $1 \ \mathrm{Myr}$, assuming  $1 \ M_\odot$ for the stellar masses, and  $1 \ M_J$ for the giant planet masses. It is evident that a significant number of JuMBOs can be produced in the ONC for wide planetary systems with closely paired giant planets, enabled by stellar flybys. In particular,
the observations { reported} by \cite{Pearson2023} revealed about 40 JuMBOs for a stellar population of { $\sim 3500$ stars \cite{Hillenbrand1998}}, hence suggesting a production efficiency of { $\sim 1\%$}
per planetary system. 
{  According to the rates per planetary systems illustrated in the upper panel of Fig.~\ref{fig:rate}, 
this scenario is plausible if there exists a sufficient number of systems with multiple giant planets, where planets orbit at distances beyond 100 AU, as predicted through gravitational instability formation \cite{Boley2010} and observed in many systems \cite{Nielsen2019,Vigan2021}. } Conversely, a smaller number of JuMBOs would indicate that either two giant planet systems are rare or that their planets are located much closer to their host stars. 

An important feature of the JuMBO formation model proposed here is its strong environmental dependence, being a dynamically-induced phenomenon. To illustrate this effect, we provide a comparative rate study for a typical open cluster in the bottom panel of Fig~\ref{fig:rate}. It is evident that even in the optimal case of two wide-orbit planets for every star, the absolute JuMBO production rate remains low. Consequently, JuMBO production via this mechanism would not be a viable explanation if such objects were to be found in low-density star clusters.}

{We note that our rate calculations for the JuMBO
{production} are broadly consistent with those computed by \citep{Portegies2023}. Their low-density model with number density $n\sim 5.4\times 10^3$~pc$^{-3}$ (lower than that in Trapezium) was found to be insufficient for generating the expected number of JuMBO in Trapezium, while their high-density model, with $n\sim 2.9\times 10^5$~pc$^{-3}$ (larger than the Trapezium one), could produce JuMBOs at a satisfactory rate.
However, in such a dense environment, subsequent ionization can become important. The ionization cross-section for a binary composed of two { Jupiter mass planets, with a SMA of $100$~AU, in a stellar environment with 1 solar mass stars and velocity dispersion $\sigma_v$ of 2 km/s, is numerically found to be roughly 5.5$\times10^5$ AU$^2$.} 
The average ionization time, $1/(n\sigma_{\rm ioin} \sigma_v)$, for  $n\sim 2.9\times 10^5$~pc$^{-3}$ is approximately 
{ $1.2\times10^5$} years, which is shorter than their simulation duration of 1~Myr. This explains the scarcity of JuMBOs observed in very high-density star clusters within their study. This rapid ionization also leads to a significant reduction in the JuMBO to FFP ratio. Consequently, a  lower JuMBO to FFP ratio was observed in their study.

\subsection*{Conclusion}\label{sec13}

Dynamical encounters in interacting stellar environments can give rise to a variety of planetary architectures which collectively would be difficult to otherwise explain via conventional models of planetary formation.
Here, via a large suite of dedicated $N$-body simulations, we have shown that close flybys in {dense}
stellar clusters {\em unavoidably} lead to a sizeable fraction of free-floating binary planets, in addition to the already known single free-floating planets.   

Our simulations have allowed us to quantify their probabilistic outcomes depending on the initial planetary properties and those of their host stellar cluster, as well as characterizing their
orbital properties. JuMBOs can be produced with a wide range of semi-major axes, largely correlated with the difference between the original orbital distances of the ejected planets.
Most notably, they are expected to have high eccentricity with a superthermal distribution, unlike in the primordial formation channel from ISM clouds, which predicts a sub-thermal distribution.

Recent observations with the {JWST} have identified some of these 
{  potential}
candidates in a very dense star cluster \cite{Pearson2023}. 
{ 
We have shown that JuMBO formation from ejections could broadly account for the reported candidates
provided that there is a sufficient number of multiple
giant planet systems with planets orbiting at distances beyond a few tens of AU.}
With much more data expected in the years to come, our results will allow to further test this dynamical formation scenario. A characterization of the JuMBO orbital properties, and their relative fraction with respect to that of FFJ,  will allow us to probe primordial planetary architectures
and thus help discriminate between competing theories of planetary formation.

\section*{Methods}\label{sec11}
\subsection*{Cross section calculation}

 To calculate the cross-sections of single free-floating Jupiter-mass planets (FFJs) and JuMBOs resulting from stellar flybys, we investigate the parameter space delineated by $v_\infty/v_2 \in [10^{-1},10]$ and $a_1/a_2\in [0.25, f_{\rm max}]$ by using the high-precision $N$-body code {\tt Spacehub} \cite{Wang2021}. We use 20 equally spaced grids for each parameter. Here, $f_{\rm max}$ represents the maximum allowable ratio that ensures
 {
\begin{equation}
a_2-a_1>R_{\rm Hill} \sim \left(\frac{2m}{3M_1}\right)^{1/3}\left(\frac{a_1+a_2}{2}\right)\,.
\end{equation}
}
This is a key condition for the two planets to remain stable { for a sufficient amount of time \cite{Chambers1996}} prior to scattering. For each grid, we conduct  $10^8$ scattering experiments, ensuring a uniform distribution for $\nu_1$, $\nu_2$, $\cos I$, $\Omega$, $\theta$. $M_2$ is generated at asymptotic infinity within a circle of radius $b_{\rm max}$, 
\begin{equation}
b_{\rm max}  = r_{\rm p,max}\sqrt{1+\frac{2G(M_1+M_2)}{v_\infty^2r_{\rm p,max}}}\,,
\end{equation}
where $r_{\rm p, max}$ is the maximum closest approach distance, which we set to be at 5$a_2$ as a conservative estimate to ensure that all FFJs and JuMBOs formed from a stellar flyby are included in the scattering outcomes.

At the end of each scattering experiment, the pairwise SMA and eccentricity between particle $m_i$ and $m_j$ are calculated based on their pairwise relative position $\mathbf{r}_{ij}$ and velocity $\mathbf{v}_{ij}$,
\begin{eqnarray}
\frac{G(m_i+m_j)}{-2a_{ij}}&=&\frac{\mathbf{v}_{ij}\cdot \mathbf{v}_{ij}}{2}- \frac{G(m_i+m_j)}{{r}_{ij} }\label{eq:a}\\
\mathbf{e}_{ij}&=&\frac{\mathbf{v}_{ij}\times(\mathbf{r}_{ij}\times\mathbf{v}_{ij})}{G(m_i+m_j)} - \frac{\mathbf{r}_{ij}}{r_{ij}}\label{eq:e}\,,
\end{eqnarray}
where $r_{ij}=|\mathbf{r}_{ij}|$. If $a_{ij}>0$, it means that particles $m_i$ and $m_j$ are in a bound state as a binary, provided they are not in a bound state with other particles. FFJs are characterized by planet-mass particles 
(denoted by $m$ here)
that are not in a bound state with any other particle. In contrast, JuMBOs are identified when two planet-size particles 
are in a bound state and not bound to any other particle. The duration of the scattering experiments ensures that the particles are well-separated before the scatterings are terminated.

The cross-sections of FFJ and JuMBO are respectively calculated via 
\begin{eqnarray}
    \sigma_{\rm FFJ} &=&\pi b_{\rm max}^2\frac{N_{\rm FFJ}}{N}\\
    \sigma_{\rm JuMBO} &=&\pi b_{\rm max}^2\frac{N_{\rm JuMBO}}{N}\,,
\end{eqnarray}
where $N$ represents the total number of scatterings, $N_{\rm FFJ}$ denotes the number of scatterings that result in FFJ production, and $N_{\rm JuMBO}$ indicates the number of scatterings leading to JuMBO production.

\subsection*{Rate estimation}
The rate of  JuMBO formation per planetary system can be estimated via the equation
\begin{equation}
\mathcal{R}_{\rm JuMBO}\sim  f_{2J} \int nv_\infty \sigma_{\rm JuMBO}(a_1, a_2, v_\infty) f(v_\infty)f(a_1, a_2)d a_1 da_2  dv_\infty\,,
\end{equation}
where $f_{2J}$ is the fraction of planetary systems which host two Jupiter-mass planets, $n$ is the number density of the stars, $f(v_\infty)$ is the velocity distribution function  normalized such that $\int f(v_\infty) dv_\infty=1$, and $f(a_1, a_2)$ is the joint semi-major axis distribution function of two Jupiter mass planets that satisfy $\int f(a_1, a_2) da_1 da_2=1$. For thermal systems, the velocity distribution is  Maxwell-Boltzmann. The joint probability functions $f(a_1, a_2)$ and $f_{2J}$ are poorly constrained both observationally due to the observation bias toward closer planets to their host star, and theoretically due to different predictions made by different planet formation theories: while the core accretion theory (e. g. \cite{Pollack1996I}) has difficulties in accounting for planets on such wide orbits, they can easily be accounted for by the disk-instability model \cite{Boss1997}.

{Nonetheless, we can make generic rate predictions based on parametrizations of these uncertain functions.
For a given initial $f_{\rm 2J}(t=0)$, we can assume $f(v_\infty)=\delta(\sigma_v)$, $f(a_1,a_2)=\delta(a_1)\delta(a_2)$ (which implies that other close encounters do not significantly modify this distribution),  and thus  estimate the absolute JuMBO production rate as a function of $a_1$ and $a_2$. Here, $\sigma_v$ represents the velocity dispersion of the star cluster, which has density $n$.  With this,  the number of JuMBOs produced per planetary system over a period $t$
can be estimated 
by using the following equation
\begin{equation}
    N_{\rm JuMBO} = \int f_{2J}(t) n \sigma_v \sigma_{\rm JuMBO} dt\,.
\end{equation}
The time-dependent function $f_{2J}(t)$ is derived from the equation
\begin{equation}
\frac{df_{2J}}{dt} \sim -\frac{f_{2J}\sigma_{\rm FFJ}\sigma_v}{V_c}-\frac{f_{2J}\sigma_{\rm JuMBO}\sigma_v}{V_c} \sim -\frac{f_{2J}\sigma_{\rm FFJ}\sigma_v}{V_c}\,,
\end{equation}
where $V_c$  represents the volume of the star cluster, and we have used the fact that $\sigma_{\rm FFJ}\gg\sigma_{\rm JuMBO}$.
This yields the solution for $f_{2J}$ as  
\begin{eqnarray}
    f_{2J} &=& f_{2J}(0) e^{-\frac{t}{\tau_{\rm ej}}}\,,\\
    \tau_{\rm ej} &=& \frac{V_c}{\sigma_{\rm FFJ}\sigma_v}\,.
\end{eqnarray}

In the optimal scenario, i.e. assuming $f_{2J}(0)=1$, we can thus derive an upper limit to the number of JuMBOs produced via ejection per planetary system over a period of time in the cluster. {Note that this assumption makes our rate computation conservative with respect to the fact that a large fraction of planetary systems has $N>2$ planets (see e.g. \citep{Miret-Roig2022}).} The results are displayed in 
Figure~\ref{fig:rate}, where we contrast the JuMBO numbers for the Trapezium cluster ($n\sim 5\times10^4 \ \mathrm{pc}^{-3}$ and $\sigma_v\sim2 \ \mathrm{km/s}$ in the inner region $< 0.2 \mathrm{pc}$ \cite{Hillenbrand1998})
with those of a more typical, less dense stellar cluster ($n\sim 10^2 \ \mathrm{pc}^{-3}$ and $\sigma_v\sim 1 \ \mathrm{km/s}$). }

The rate of the single FFJs can be similarly calculated as
\begin{equation}
\mathcal{R}_{\rm FFJ}\sim \sum_i f_{iJ} \int nv_\infty \sigma_{\rm FFJ}(a_1,...,a_i, v_\infty) f(v_\infty) dv_\infty f(a_1,...a_i)d a_1...d a_i\;,
\end{equation}
where $f_{iJ}$ is the fraction of planetary systems hosting $i$ Jupiter-mass planets, and $\sigma_{\rm FFJ}(a_1,...,a_i, v_\infty)$ represents the overall
cross section of FFJ production for planetary systems with $i$ Jupiter-mass planets. This cross section takes into account the 
cases of 
multiple single ejections, albeit 
it is dominated by the single ejection
of the outermost planet. Observationally, $f_{1J}$ predominates over $f_{2J}$, while $f_{3J}$ and subsequent fractions are negligible.

 The relative rate of JuMBO to FFJ is then given by  the ratio $\mathcal{R}_{\rm FFJ}/\mathcal{R}_{\rm JuMBO}$, which can be approximated as 
 \begin{equation}
 \frac{\mathcal{R}_{\rm JuMBO}}{\mathcal{R}_{\rm FFJ}}\sim \frac{f_{2J}}{f_{1J}+f_{2J}}\frac{\sigma_{\rm JuMBO}}{\sigma_{\rm FFJ}}\,.
 \label{eq:relrate}
 \end{equation}

\subsection*{Viability of the model initial conditions in a dense star cluster}

Our model relies on the assumption that
the conditions governing planet formation in a densely populated star cluster, such as Trapezium in the Orion nebula, remain largely unaffected
by the high stellar density. We will discuss the validity of our assumption in the following.

The first issue we address concerns disk photoevaporation induced by the intense radiation in densely clustered environments, which is a well-recognized factor in planetary formation (see e.g. \citep{Adams2010} for a discussion and review of this topic). 

In terms of theoretical considerations, we refer to Section 5 of \citep{Adams2010}.
Figure 3 illustrates the photoevaporation timescales for a solar-like system,
presenting the dependency on radius for varying magnitudes of the photoionizing flux.
For a dense cluster with $\sim 2000$ stars, the ioionizing flux
is $\sim 3\times 10^{11}-3\times 10^{12}$~photons~s$^{-1}$~cm$^{-2}$.
At a distance $r\sim 150$~AU, the corresponding evaporation time is $ \sim 10$~Myr. This timescale is much larger than the one for giant planet formation at large radii in the gravitational instability model, which is $< 10^5$~yr \citep{Boley2009} at large radii.
Therefore, from a theoretical point of view, the conditions in the Trapezium cluster do not preclude giant planet formation on a $\gtrsim 100$~AU scale.

Additionally, and perhaps most importantly, there have been several direct observations of the star forming region in the Trapezium
cluster of the Orion nebula. Notably, \citep{Vicente2005} found that approximately 40\% of circumstellar disks in the Trapezium cluster
possess radii exceeding 50 AU, with several extending into the 200-300~AU range.  Also interestingly from an observational perspectives are
the data collected by \citep{Mann2009}. They
showed that the fraction of disks that contain a minimum mass solar nebula within 60 AU of the Trapezium cluster is comparable to that in the Taurus region,
which has no high mass stars and very little radiation background, hence further confirming that the background ionization flux does not
play a relevant role for the size of disks in the Trapezium cluster.

The second issue we address in relation to the viability of the initial conditions of our model is whether the relatively high rate of strong interactions 
could impact the circumstances conducive to planetary formation. More specifically, 
if interactions are too frequent during the planet formation phase, they may influence the likelihood of forming outer giant planets, and hence affect the required initial conditions for our model. For the stellar density in the Trapezium cluster, we estimated (Sec.4.2) the rate of strong encounters leading to a JuMBO ejection to be of about 0.1-0.5 per Myr per planet (cfr. Fig.\ref{fig:rate}). The rate of encounters leading to FFJs is higher, up to a few per Myr. However, as long as the timescale for strong encounters leading to ejections does not exceed the timescale for planetary formation, which, for gravitational instability at large radii is $< 10^5$~yr \citep{Boley2009}, we can safely assume that giants can be still formed and subsequently ejected.

\backmatter


\subsection*{Data availability}
\noindent

\noindent
Data are available at \url{https://figshare.com/articles/dataset/jumbo_dataset_1/25331110}.

\subsection*{Code availability}
\noindent
The code {\tt Spacehub} is available at \url{https://github.com/YihanWangAstro/SpaceHub}.
The problem generator and data process script are available at \url{https://github.com/YihanWangAstro/JuMBO-code/tree/main}.

\section*{Acknowledgements}
R.P. acknowledges support by NSF award AST-2006839. Y.W. and Z.Z. acknowledge support from NASA 80NSSC23M0104 and the Nevada Center for Astrophysics. Y.W. acknowledges the very useful discussion with Douglas N.C. Lin.

\section*{Author Contributions} 
R.P. proposed the model idea;  Y.W. devised the numerical experiments, processed the data, and performed the calculations. Z.Z. provided input on models of planetary formation. R.P. and Y.W. drafted the first version of the manuscript. All authors contributed to the analysis and interpretation of the data, as well as to the final version of the manuscript.

\section*{Competing Interests} 
The authors declare no competing interests.

\section*{Tables}

\section*{Figure Legends/Captions }\label{sec6}

\begin{figure}[H]
    \includegraphics[width=\columnwidth]{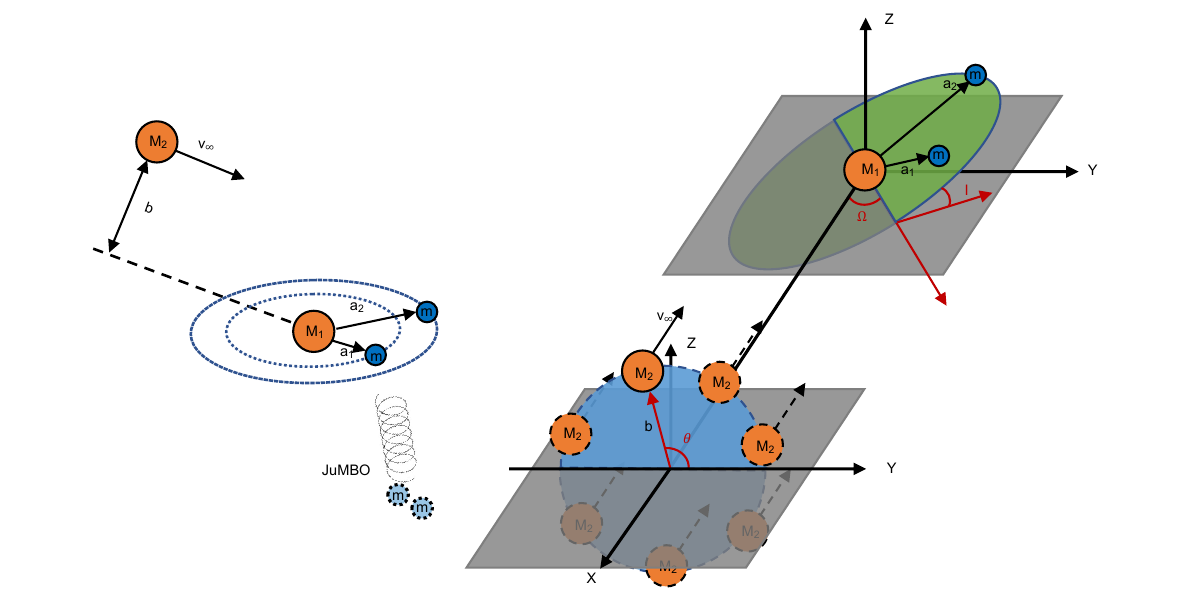}
    \caption{\textbf{Schematics of JuMBO production from stellar encounters.} {\em Left}: Schematics of the astrophysical scenario we explore. A close stellar flyby 
    to a planetary system results in the ejection of two planets, which thereof remain bound, forming a floating planetary binary.
    {\em Right: }Schematics of the scattering experiments set to explore the occurrence of such a scenario.    
    Two equal-mass, co-planar planets orbit a star of mass $M_1$. An interloper star of mass $M_2$ flies by 
    with asymptotic velocity $v_\infty$ parallel to the $X$ direction, impact parameter $b$ and angle $\theta$ in the plane perpendicular to the direction of motion of $M_2$.
    The planetary orbital plane forms an angle  $I$, and is rotated by an angle $\Omega$, with respect to the direction of motion of $M_2$.}
    \label{fig:schematics1}
\end{figure}

\begin{figure}[H]
    \includegraphics[width=\columnwidth]{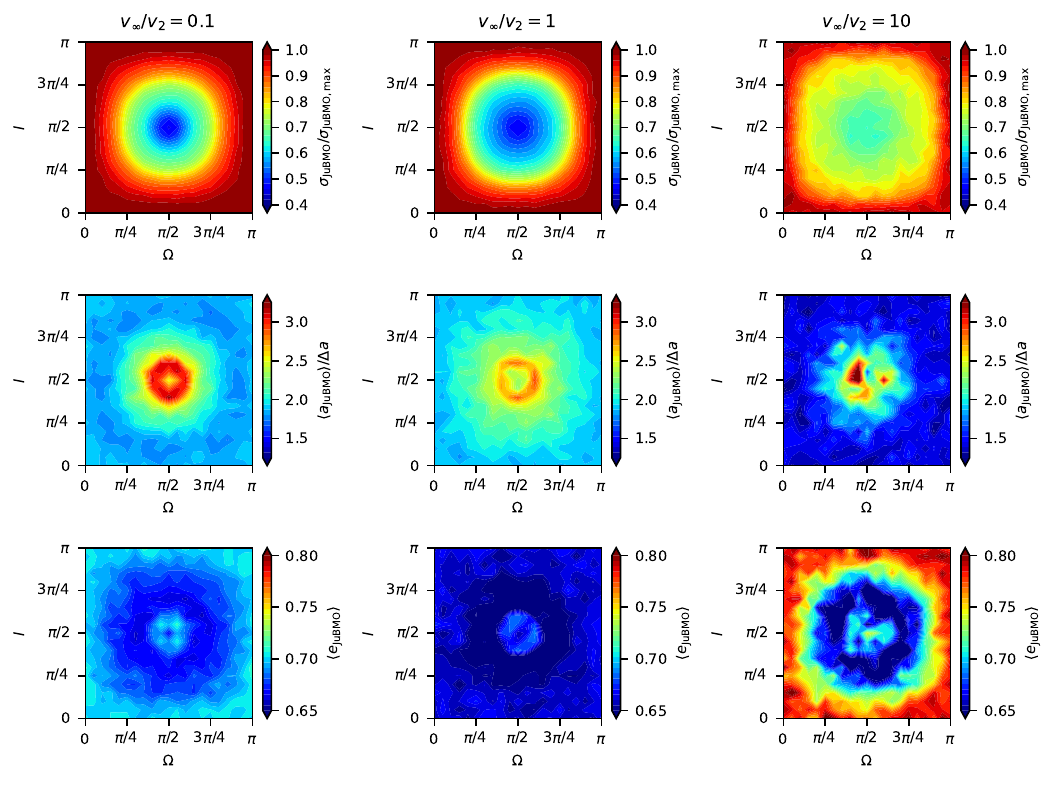}
    \caption{\textbf{Angular dependency on JuMBO production cross section, semi-major axis, and eccentricity.} Dependence on the geometry of the encounter
    of the differential cross section
    for JuMBO production (top panels), of the SMA of the JuMBO (middle panels), and their eccentricity (bottom panels).
    From left to right, the velocity of the scatterer is increasing.
    In all the cases, the initial ratio between the SMA of the two planets is $a_1/a_2=0.7$ while the angle  
    $\theta\in[0,2\pi]$.}
    \label{fig:Omega-I}
\end{figure}

\begin{figure}[H]
    \includegraphics[width=\columnwidth]{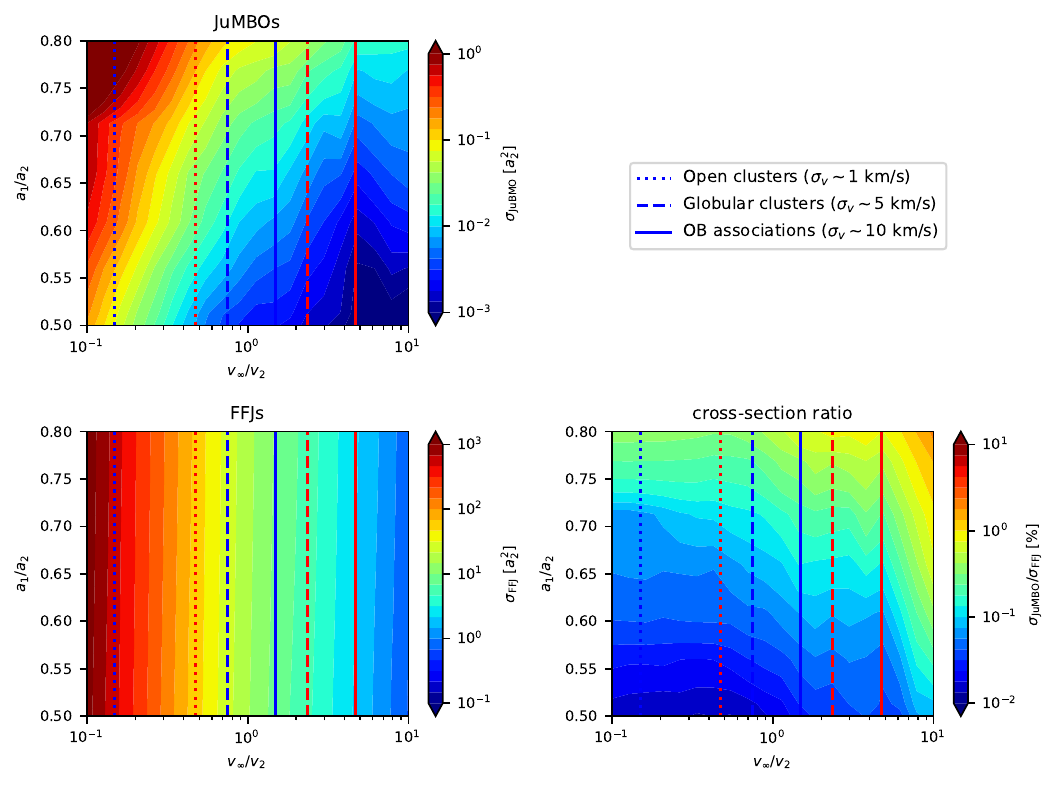}
    \caption{\textbf{Cross section for the production of JuMBOs  and FFJs.} The relative cross section of JuMBO to single-floating planets 
    is displayed in the bottom right panel. 
    The vertical lines in all the panels show the value of $v_\infty/v_2$ corresponding to $a_2=10$~AU (blue lines) and $a_2=100$~AU (red lines), each for the three representative dispersion velocities of the star cluster indicated in the inset of the top panel.    
    }
    \label{fig:cross-sections}
\end{figure}

\begin{figure}[H]
    \includegraphics[width=\columnwidth]{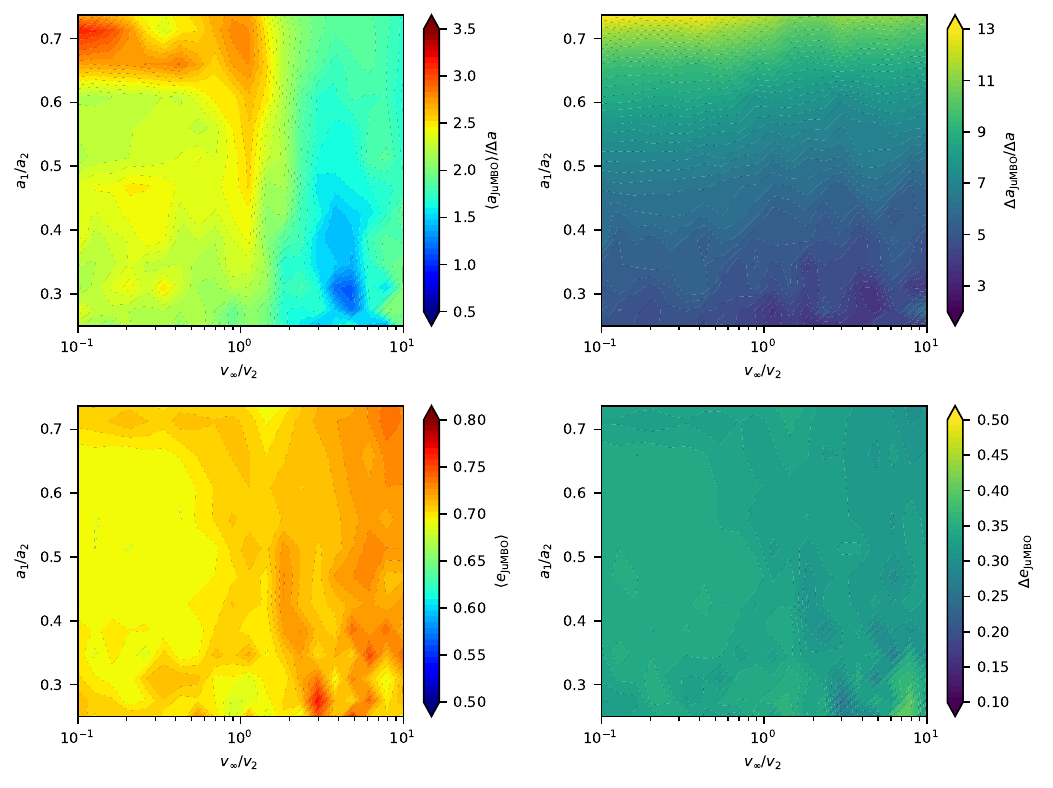}
    \caption{\textbf{Orbital parameters of the ejected JuMBO.} The top panels display the mean (left panel) and the standard deviation (right panel) of the
semi-major axis of the binary while the bottom panels correspondingly show  the same quantities but for the eccentricity. 
JuMBO are expected to be eccentric and with semi-major axis generally larger than
 $ \Delta a$, dependent on
the initial $a_2/a_1$.}
    \label{fig:jumbo-ae}
\end{figure}

\begin{figure}[H]
    \includegraphics[width=\columnwidth]{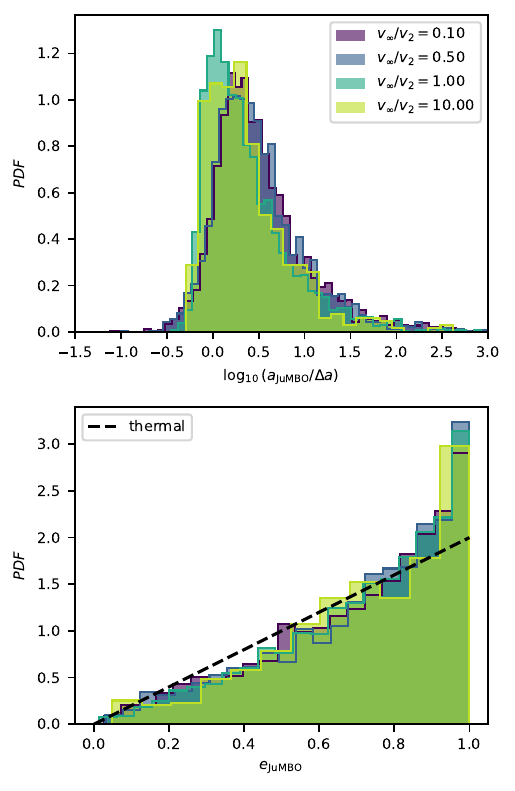}
    \caption{\textbf{The distribution of semi-major axis and eccentricity for JuMBO produced from stellar flybys.} The initial semi-major axis ratio $a_1/a_2$ is fixed at 0.7 for all values of $v_\infty/v_2$. For this ratio, at which the JuMBO cross-section is higher, the semi-major axis (upper panel) is broadly centered around $\sim 3 (a_2-a_1)$. The eccentricity (bottom panel) is superthermal. Such eccentricity distribution is a distinctive feature of this formation mechanism, which can observationally distinguish it from primordial formation in ISM clouds.}
    \label{fig:dist-ae}
\end{figure}

\begin{figure}[H]
    \includegraphics[width=\columnwidth]{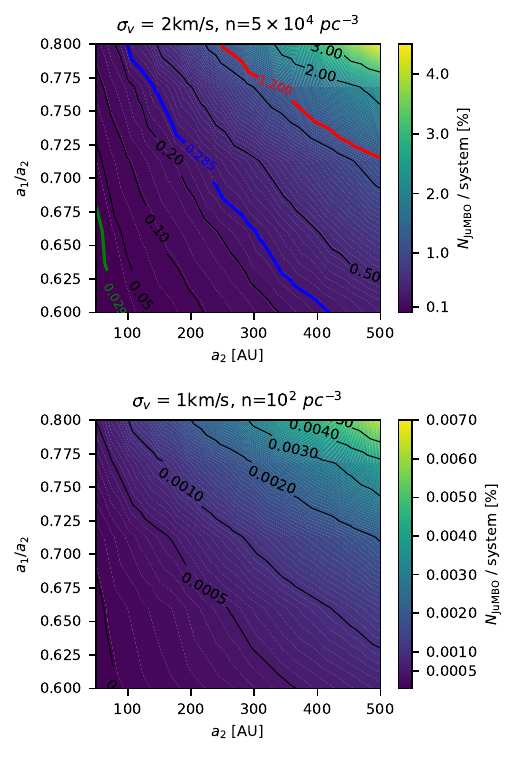}
    \caption{\textbf{The upper limit for the number of JuMBOs produced per planetary system over $1 \ \mathrm{Myr}$ due to stellar ejections.} \textit{Upper panel:} stellar density and velocity dispersion as measured in the Trapezium cluster (adopted from \cite{Hillenbrand1998}).
    { The green, blue and red lines  mark the regions (to their right)  required to produce $>1, >10$ and $>42$ JuMBOs in the Trapezium cluster, respectively.}
    \textit{Bottom panel:} conditions typical of an open cluster. }
    
    \label{fig:rate}
\end{figure}

\bibliography{sn-bibliography}

\end{document}